\documentclass{czjphys}         
\newcommand{\beq}{\begin{equation}} 
\newcommand{\eeq}{\end{equation}} 
\newcommand{\bx}[1]{\mbox{\boldmath $#1$}} 
\begin{document}
\title{Diquark model for $J/ \psi$ baryonic decays}  
\authori{A.~Ciepl\'y}      
\addressi{Institute of High Energy Physics, P.O.Box 918, Beijing
100039, China 
\\ Nuclear Physics Institute, 25068 \v{R}e\v{z}, Czech Republic}
\authorii{B.S.~Zou}     
\addressii{Institute of High Energy Physics, P.O.Box 918, Beijing
100039, China 
\\ CCAST (World Laboratory), P.O.Box 8730, Beijing 100080, China}
\authoriii{}    \addressiii{}
\authoriv{}     \addressiv{}
\authorv{}      \addressv{}
\authorvi{}     \addressvi{}
%
\headauthor{A.~Ciepl\'y, B.S.~Zou}            
\headtitle{Diquark model for $J/ \psi$ baryonic decays}             
\lastevenhead{A.~Ciepl\'y, B.S.~Zou} 
\refnum{A}
\daterec{XXX}    
\issuenumber{0}  \year{2001}
\setcounter{page}{1}
\maketitle

\begin{abstract}
The baryonic decays of $J/ \psi$ provide a new way to study the
internal structure of baryons. We apply a simple diquark model to
the calculation of the decay cross-sections for the reactions $J/
\psi \longrightarrow \bar{p}p$, $\bar pN^*(1440)$,
$\bar{N}^{\ast}N^{\ast}$, $\bar{\Lambda}\Lambda$ and
$\bar\Sigma^0\Sigma^0$. The results are different from those given
by the ordinary constituent quark model. Hence these reactions may
provide a new check of two different pictures for the baryons.
\end{abstract}

\section{Introduction}
\label{sec:int}

In spite of a long history of baryon spectroscopy there are still
many questions without clear answers. We know that the baryons are
composed of three valence quarks, sea quarks and gluons but we are
not certain if the valence quarks have constituent or current
nature, if they cluster into diquarks or are well separated from
each other, whether the sea quarks are in the form of $\bar{q}q$
soup or mesons etc. For a long time we have also missed direct
sources of information on the properties of nucleon excitation
states and could build our knowledge about them almost entirely on
results from partial wave analysis of $\pi N$ scattering data. The
situation is changing dramatically with new experimental data
coming from facilities such as CEBAF at JLAB, ELSA in Bonn, GRAAL
in Grenoble or from BEPC in Beijing.

A long-standing problem in $N^{\ast}$ physics is about the nature
of the Roper resonance $N^{\ast}(1440)$ which is considered to be
the first radial excitation state of the nucleon. However, various
quark models have difficulty to explain its mass and
electromagnetic coupling, so it was suggested that it may be a
gluonic excitation of the nucleon, a hybrid baryon. An ideal tool
for studying the properties of $N^{\ast}$ resonances are the
decays of $J/ \psi \rightarrow \bar{p}p \pi^{0}$ and $J/ \psi
\rightarrow \bar{p}p \pi^{+} \pi^{-}$  since in these processes
the $\pi^{0} p$ and $p \pi^{+} \pi^{-}$ systems are limited to
isospin $1/2$ states. The $J/ \psi$ decays into baryon-antibaryon
states discussed in the present paper provide another way to probe
the internal structure of baryons.

The $J/ \psi$ decay cross sections for $\bar{p}p$, $\bar{p}N^{*}$
$\bar{N^{\ast}}N^{\ast}$, $\bar\Lambda\Lambda$ and
$\bar\Sigma^0\Sigma^0$ final states were calculated in
Ref.\cite{ZPP01,Pingrg} by using a simple quark model. In the
present paper we use the diquark model and look into the
possibility of forming the final state baryons from the $\bar{q}q$
(generated in $J/ \psi$ decay) and diquark-antidiquark
($\bar{D}D$) pairs (created as vacuum excitation). The model
resembles the standard quark-pair-creation model \cite{YOP73} used
extensively to describe the mechanism of hadronic decays. Our
diquark model calculation gives a different prediction for these
processes which can be compared with the previous calculation by
the quark model\cite{ZPP01,Pingrg}. Future experimental data on these
reactions may provide a check on these two different pictures for
baryons.

\section{Diquark model formalism}
\label{sec:diq}

We assume that the decay of $J/ \psi$ meson leads to a creation of
$q \bar{q}$ pair and the final state baryons are formed by their
coupling to the diquark-antidiquark pair produced with the vacuum
quantum numbers $J^{PC}=0^{++}$. Restricting ourselves to the
SU(3) flavour sector we have the following diquark nomenclature
\cite{VW-91}
$$
\begin{array}{rcll}
(qq') & \sim & (ud-du) & \;\;\;\hbox{\rm scalar-isoscalar diquark} \\
(qs)  & \sim & (us-su) & \;\;\;\hbox{\rm scalar-isodublet diquark} \\
\left[qq'\right] & \sim & uu      & \;\;\;\hbox{\rm vector-isotriplet diquark} \\
\left[qs\right]  & \sim & (us+su) & \;\;\;\hbox{\rm vector-isodublet diquark} \\
\left[ss\right]  & \sim & ss      & \;\;\;\hbox{\rm vector-isosinglet diquark,}
\end{array}
$$
in which $q$ (or $q'$) stands for either the {\it up} or {\it down} quark and
we use the round and box brackets for the axial scalar and the axial
vector diquarks, respectively. The effective diquark-antidiquark
vacuum creation amplitude can be written as
\beq
\langle \bar{D}D \mid V_{\rm eff} \mid 0 \rangle =
f_{\bar DD} \, (-1)^{T_{\bar D}-M_{\bar D}}
      (T_{\bar D} M_{\bar D} T_D M_D | 0 0)\:
      \eta \;\;\;\; .
\eeq A similar structure was adopted in Ref.~\cite{CLZ93} for the
$\bar{D}D \longrightarrow \bar qq$ process. The coupling parameter
$f_{\bar DD}$ may be different for the various $\bar{D}D$ species
listed above. However, for the sake of simplicity we shall assume
that this parameter does not depend on the diquark type. The phase
$\eta=1$ stands for the creation of scalar diquarks and
$\eta=(-1)^{1-\epsilon_{\bar D}}\delta_{\epsilon_{\bar
D}-\epsilon_D}$ corresponds to the scalar coupling of vector
diquark (with polarization $\epsilon_D$) and vector antidiquark
(with polarization $\epsilon_{\bar D}$).

The intermediate $\bar{q}q$ pair produced in $J/ \psi$ decay can
be either in the $^{3}S_{1}$ or in the $^{3}D_{1}$ state. In the
$J/ \psi$ rest system and for the initial $J_{z}=J=1$ polarization
the relative $\bar{q}q$ wave function can be written as \beq
\langle \bx{p} \mid \bar{q}q, J=1 \: J_{z}=1  \rangle = \sum_{LM}
c_{LM} \phi_{L}(p) Y_{LM}(\bx{\hat{p}}) \mid \bar{q}q, \:(1-M)
\rangle \eeq where $\mid \bar{q}q, \:\mu \rangle$ stands for the
spin-flavour part of the $\bar{q}q$ wave functions and $\bx{p}$
denotes the relative $\bar{q}q$ momentum. The radial wave
functions $\phi_{L}(p)$ and the coefficients $c_{LM}$ are
normalized as
$$
\begin{array}{l}
\sum_{L} \int_{0}^{\infty} dp\: \phi_{L}^{2}(p) = 1\; , \;\;\;\;
\sum_{M} c_{LM}^{2} = 1 \\
(c_{00}=1, \; c_{20}=\sqrt{1/10}, \; c_{21}=-\sqrt{3/10}, \; c_{22}=\sqrt{3/5}).
\end{array}
$$
The $\bar{q}q$ pair couples with the $\bar{D}D$ pair forming the final state
baryon $B$ and antibaryon $\bar{B}$. The spin-flavour overlap amplitudes
follow from the SU(3) quark-diquark decomposition of the baryonic
states (see Appendix). It is easy to show that
\beq
\langle \bar{B}B, \:\mu' \mid V_{\rm eff} \mid \bar{q}q, \:\mu \rangle
 =  V_{q}(B)\: \delta_{\mu'\mu} \;\;\;\; ,
\label{eq:Vq}
\eeq
where we introduced the $\bar{B}B$ spin triplet states in the same fashion as
the $\bar{q}q$ spin-flavour states. The amplitudes (\ref{eq:Vq}) depend on
both the baryon specification as well as on the flavour of the intermediate
quark. For a reference we give the amplitudes relevant for the considered
processes in Table~\ref{tab:Vq}. It is worth noting that the amplitudes
(\ref{eq:Vq}) vanish if they are derived for the spin-singlet $\bar{B}B$
states in accordance with the required parity conservation.

%
%
\bt[t]                          
\label{tab:Vq}
\caption{The spin-flavour overlap amplitudes [see Eq.(\ref{eq:Vq})] derived for
various $\bar{B}B$ final states. All coefficients $V_{q}(B)$ and the summed amplitudes
$\sum f_{q}V_{q}$ are presented in units of $f_{\bar{D}D}/2$.}
\vspace{2mm}
\small
\bc                             
\begin{tabular}{ccccc}
\hline\hline
$\bar{B}B$ state & $V_{u}(B)$ & $V_{d}(B)$ & $V_{s}(B)$ & $\sum f_{q}V_{q}$ \\
\hline
 $\bar{p}p$,$\bar pN^*$,$\bar N^*N^*$ & $1-1/(9\sqrt{3})$
 & $-2/(9\sqrt{3})$ & $0$ & $[1-1/(3\sqrt{3})]f_{u}$ \\
 $\bar{\Lambda}\Lambda$ & $0$ & $0$ & $2/3$ & $2f_{s}/3$ \\
 $\bar{\Sigma^{0}}\Sigma^{0}$ & $4/(9\sqrt{2})$ & $4/(9\sqrt{2})$
    & $-2/(9\sqrt{3})$ & $8f_{u}/(9\sqrt{2})-2f_{s}/(9\sqrt{3}) $ \\
    \hline\hline
\end{tabular}
\vspace{-1mm}
\ec                             
\et                             

The complete transition matrix elements for charmonium decay into $\bar{B}B$
states (with spin projection $\mu$) can be written as
\beq
M_{\mu}=\sum_{q,LM} f_{q} V_{q}(B)\: c_{LM} I_{\rm space}^{LM} \delta_{\mu (1-M)}
\label{eq:Mmu}
\eeq
where the parameter $f_{q}$ represents the amplitude of creating the specific
$\bar{q}q$ pair. The space integral reads
\beq
I_{\rm space}^{LM}=\frac{1}{8} \int d^{3}k\: \phi_{L}(k)\: Y_{LM}(\bx{\hat{k}})\:
\phi^{\ast}_{B}(\bx{B}-\frac{3}{2}\bx{k})\:
\phi^{\ast}_{\bar{B}}(\frac{3}{2}\bx{k}-\bx{B})
\label{eq:space}
\eeq
with $\bx{B}$ standing for the final state baryon momentum and $\phi_{B}$ denoting
the intrinsic spatial wave function of the baryon. We have assumed a simple gaussian
spatial distribution for both the intermediate $\bar{q}q$ state and for the baryon
intrinsic wave function. Only the $L=0$ and $L=2$ partial waves contribute
to the $\bar{q}q$ state and we have used
\begin{eqnarray}
\phi_{0}(x)& = & n_{0} \sqrt{4\pi} \left( \frac{1}{\alpha \pi^{1/2}} \right)^{3/2}
\exp (-\frac{x^{2}}{8\alpha^{2}}) \nonumber \\[-3mm]
 & & \label{eq:phiQ} \\[-3mm]
\phi_{2}(x)& = & n_{2} \sqrt{4\pi} \left( \frac{1}{\alpha \pi^{1/2}} \right)^{3/2}
\sqrt{\frac{1}{60}} \frac{x^{2}}{\alpha^{2}}\: \exp (-\frac{x^{2}}{8\alpha^{2}})
 \nonumber
\end{eqnarray}
where the factors $n_{L}$ must satisfy the normalization condition
$\sum_{L} n_{L}^{2}=1$. As there is no evidence that the strength of
$\bar{q}q$ formation process depends on $L$ we simply assume
$n_{0}=n_{2}=\sqrt{1/2}$ in our calculation. The harmonic oscillator
eigenfunctions are used for the baryons in their center-of-mass system, e.g.
\begin{eqnarray}
\phi_{\rm p}(x)& = & \sqrt{4\pi}\: \left( \frac{1}{\beta \pi^{1/2}} \right)^{3/2}
\exp (-\frac{x^{2}}{12\beta^{2}}) \nonumber \\[-3mm]
 & & \label{eq:phiB} \\[-3mm]
\phi_{N^{\ast}}(x)& = & \sqrt{4\pi}\: \sqrt{\frac{12}{5}}
\left( \frac{1}{\beta \pi^{1/2}} \right)^{3/2}
\left( 1-\frac{x^{2}}{18\beta^{2}} \right) \exp (-\frac{x^{2}}{12\beta^{2}})
\nonumber
\end{eqnarray}
for the proton and the $N^{\ast}(1440)$ resonance. The momentum $\bx{x}$ stands for
the relative $\bar{q}q$ momenta and for the quark-diquark momenta in Eqs.(\ref{eq:phiQ})
and (\ref{eq:phiB}), respectively. The harmonic oscillator parameters $\alpha$ and $\beta$
characterize the size of the relevant interaction and we varied them to fit
the experimental data.

It was already noted in Ref.\cite{ZPP01} that relativistic
description seems to be more appropriate for the spatial
distribution of the final state baryon clusters. Then the integral
(\ref{eq:space}) includes the jacobian due to Lorentz
transformation from the baryon CMS to the laboratory ($J/ \psi$ at
rest) system and the internal quark-diquark momenta are
transformed appropriately as well (see \cite{ZPP01,Pingrg} for
more details). We have performed the calculation for both the
nonrelativistic as well as the relativistic baryon wave functions.

The space integrals (\ref{eq:space}) were computed numerically by
using the program RIWIAD from CERN Program Library. The decay
cross-section for $J/ \psi \longrightarrow \bar{B}B$ was then
constructed from the amplitudes $M_{\mu}$ as \beq \frac{d\Gamma
(J/ \psi \rightarrow \bar{B}B)}{d\Omega} = \frac{1}{32\pi^{2}}
\frac{\mid \bx{B} \mid}{M_{\psi}^{2}} \sum_{\mu} \mid M_{\mu} \mid
^{2} \eeq with $\Omega$ denoting the solid angle of $\bx{B}$ and
$M_{\psi}$ standing for the charmonium mass.

\section{Results and discussion}
\label{sec:res}

There are three parameters in our calculation. The
harmonic-oscillator parameters $\alpha$ and $\beta$ determine the
shape of the angular distribution \beq \frac{d\Gamma(J/ \psi
\rightarrow \bar{B}B)}{d\Omega} = N_{\bar{B}B}\, (1+a_{B}\cos^{2}
\theta )\;\;\;\; , \eeq i.e.\ the computed parameter $a_{B}$. The
constant $N_{\bar{p}p}$ is directly related to the experimental
branching ratio of $J/ \psi \rightarrow \bar{p}p$ and can be used
to fix the overall normalization of the computed rates. Finally,
we define our third parameter as the rate $g=f_{s}/f_{u}$ and
naturally assume $f_{d}=f_{u}$. Clearly, the rate $g$ affects the
relative numbers of strange and nonstrange baryons produced in $J/
\psi$ decay. As the creation of $\bar{q}q$ pair in charmonium
decay is flavor blind one may simply assume $g=1$. However, the
variation of $g$ also accounts effectively for some differences
between creation of various $\bar{D}D$ species, i.e.\ between
different amplitudes $f_{\bar{D}D}$. In the calculation we adopted
the value $g=0.9$ suppressing partly the formation of strange baryons
in the final state. Our assumptions and the structure of the
amplitudes $V_{q}(B)$ allow to sum over $q$ in Eq.(\ref{eq:Mmu}).
The summed amplitudes $\sum_{q}f_{q}V_{q}(B)$ are given in the
last column of the Table \ref{tab:Vq}.

The results of our calculation are presented in the Table
\ref{tab:rate} for both the nonrelativistic and the relativistic
approaches. We show the relative decay rates $\Gamma_{\bar{B}B}/
\Gamma_{\bar{p}p}$, where $\Gamma_{\bar{B}B}=4\pi
N_{\bar{B}B}(1+a_{B}/3)$, and the angular distribution
coefficients $a_{B}$ in comparison with the available experimental
data. As the reliable data are limited only to the $\bar{p}p$
channel we futher assumed $\alpha=\beta$ and made an one parameter
fit to the measured value of $a_{p}$. The results shown in the
Table \ref{tab:rate} were obtained for $\alpha=\beta=0.4$ GeV and
for $\alpha=\beta=0.22$ GeV in the nonrelativistic and the
relativistic cases, respectively. Although the assumption of using
the same size parameters for both the intermediate $\bar{q}q$
state and for the space distribution of baryon clusters may not be
sound the fitted values compare well with those used in other
quark models \cite{ZPP01, ABS96}.

%
%
\bt[t]                          
\label{tab:rate}
\caption{The computed characteristics $\Gamma_{\bar{B}B}/
\Gamma_{\bar{p}p}$ and $a_{B}$ of the $J/ \psi \rightarrow
\bar{B}B$ decay rates.} 
\vspace{2mm}
\small
\bc                             
\begin{tabular}{ccccccc}
\hline\hline
 & \multicolumn{2}{c}{nonrelativistic case} & \multicolumn{2}{c}{relativistic case}
 & \multicolumn{2}{c}{experiment} \\
$\bar{B}B$ state & $\Gamma_{\bar{B}B}/\Gamma_{\bar{p}p}$ & $a_{B}$
& $\Gamma_{\bar{B}B}/\Gamma_{\bar{p}p}$ & $a_{B}$
& $\Gamma_{\bar{B}B}/\Gamma_{\bar{p}p}$ & $a_{B}$ \\
\hline
 $\bar{p}p$ & $1.00$ & $0.64$ & $1.00$ & $0.60$ & $1.00$ & $0.61(11)$\cite{DM2} \\
 $\bar{p}N^{\ast}$ & $0.90$ & $0.53$ & $0.76$ & $0.68$ & $---$ & $---$  \\
 $\bar{N^{\ast}}N^{\ast}$ & $1.05$ & $0.18$ & $0.99$ & $0.32$ & $---$ & $---$  \\
 $\bar{\Lambda}\Lambda$ & $0.56$ & $0.50$ & $0.54$ & $0.54$ & $0.61(9)$\cite{PDG} & $0.62(22)$\cite{DM2}    \\
 $\bar{\Sigma^{0}}\Sigma^{0}$ & $0.41$ & $0.50$ & $0.39$ & $0.56$ & $0.60(11)$\cite{PDG} & $0.22(31)$\cite{DM2}  \\
\hline\hline
\end{tabular}
\vspace{-1mm}
\ec                             
\et                             

The results are compatible with available experimental data on
$\bar pp$, $\bar\Lambda\Lambda$ and $\bar\Sigma^0\Sigma^0$
channels within two standard deviations. However the predicted
results for the $\bar pN^*(1440)$ and $\bar N^*N^*$ are quite
different from those given by the simple quark model \cite{ZPP01}
which predicts rates for these two channels that are at least two
times larger than the values presented here. Future experimental
results on these two channels will be helpful for examining
various model predictions and to improve our understanding of
internal quark structure of these baryons. If the experimental
$N^{*}$ production rates turn out to be much lower than the quark
(and diquark) model predictions a large component of $\pi N$ could
contribute to the $N^{*}$ internal structure. On the other hand,
too high production rates would indicate some other mechanisms
playing a role.

\bigskip
{\small This research was partially supported by the ASCR grant
IAA1048305, CAS
Knowledge Innovation Project (KJCX2-SW-N02) and the National
Natural Science Foundation of China. A.C. acknowledges the
hospitality of the IHEP in Beijing.}
\bigskip

\section*{Appendix}

In this section we show the quark-diquark wave functions of
baryons used in the present work. Suppressing color
anti-symmetrization factors the spin-up and spin-down proton
states read

\beq \mid p_{\pm}\rangle = \frac{1}{3\sqrt{2}} \left\{\pm
[ud]_{0}u_{\pm} \mp \sqrt{2} [uu]_{0}d_{\pm} \mp \sqrt{2}
[ud]_{\pm 1} u_{\mp} \pm 2[uu]_{\pm 1} d_{\mp} \right\}
                  +\frac{1}{\sqrt{2}} (ud) u_{\pm}  \;\;\;\; .
\eeq

The $N^{\ast}$ resonance has  exactly the same spin-flavour
structure and we can write for the $\Lambda$ and $\Sigma^{0}$:
\begin{eqnarray}
\mid \Lambda_{\pm}\rangle & = &
\frac{1}{2\sqrt{3}} \left\{ \pm[ds]_{0}u_{\pm}
                  \mp [us]_{0}d_{\pm} \mp \sqrt{2} [ds]_{\pm 1} u_{\mp}
                  \pm \sqrt{2}[us]_{\pm 1} d_{\mp} + \right. \nonumber \\
& + & \left. (ds)u_{\pm} + (us)d_{\pm} - 2(ud)s_{\pm}\right\}
\end{eqnarray}
\begin{eqnarray}
\mid \Sigma^{0}_{\pm}\rangle & = & \frac{1}{6} \left\{ \pm 2[ud]_{0}s_{\pm}
                  \mp [us]_{0}d_{\pm} \mp [ds]_{0} u_{\pm}
                  \mp 2\sqrt{2}[ud]_{\pm 1} s_{\mp}
                  \pm \sqrt{2}[us]_{\pm 1} d_{\mp} \pm \right. \nonumber \\
        & \pm & \left. \sqrt{2}[ds]_{\pm 1} u_{\mp} \right\}-
      \frac{1}{2}\left\{ (us)d_{\pm} + (ds)u_{\pm} \right\}
\end{eqnarray}

\end {document}